\documentclass[reprint,superscriptaddress, nofootinbib, amsmath,amssymb,aps,pra]{revtex4-2}

\pdfoutput=1

\usepackage[utf8]{inputenc}
\usepackage{CJKutf8}
\usepackage{amsthm}
\usepackage{mathrsfs}
\usepackage{physics}
\usepackage{bbm}
\usepackage{bm}
\usepackage{quantikz}
\usepackage[ruled,vlined]{algorithm2e}
\usepackage{graphicx}
\usepackage{dcolumn}
\usepackage{hyperref}
\usepackage{xcolor,soul}
\usepackage[english]{babel}
\usepackage{romannum}
\usepackage{natbib}
\usepackage{enumitem}

\usepackage{graphicx,epsfig}
\usepackage{amsfonts}
\usepackage{braket}
\usepackage{bbm}
\usepackage{algorithm2e} 
\usepackage{hyperref}
\RestyleAlgo{ruled}

\newcommand{\D}{\textnormal{D}}
\DeclareMathOperator*{\argmin}{\arg\!\min}

\DeclareMathOperator{\Ex}{\mathbb{E}}

\DeclareMathOperator{\polylog}{polylog}

\DeclareMathOperator{\regret}{Rgrt}

\newtheorem{theorem}{Theorem}
\def\<{\langle}
\def\>{\rangle}
\newcommand{\id}{\mathbbm{1}} 

\newcommand{\m}{\textemdash}

\newcommand{\dissipation}{W_{\textnormal{diss}}}

\begin{document}

\begin{CJK*}{UTF8}{gbsn}

\preprint{APS/123-QED}

\title{Quantum state-agnostic work extraction (almost) without dissipation}

\author{Josep Lumbreras}
\email{josep.lumbreras@u.nus.edu}
\affiliation{Centre for Quantum Technologies, National University of Singapore, 3 Science Drive 2, Singapore}

\author{Ruo Cheng Huang}
\email{ruocheng001@e.ntu.edu.sg}
\affiliation{Nanyang Quantum Hub, School of Physical and Mathematical Sciences, Nanyang Technological University, Singapore}%

\author{Yanglin Hu (胡杨林)}
\email{yanglin.hu@u.nus.edu}
\affiliation{Centre for Quantum Technologies, National University of Singapore, 3 Science Drive 2, Singapore}

\author{Mile Gu}
\email{mgu@quantumcomplexity.org}
\affiliation{Centre for Quantum Technologies, National University of Singapore, 3 Science Drive 2, Singapore}
\affiliation{Nanyang Quantum Hub, School of Physical and Mathematical Sciences, Nanyang Technological University, Singapore}
\affiliation{MajuLab, CNRS-UNS-NUS-NTU International Joint Research Unit, UMI 3654, 117543, Singapore}

\author{Marco Tomamichel}%
\email{marco.tomamichel@nus.edu.sg}
\affiliation{Centre for Quantum Technologies, National University of Singapore, 3 Science Drive 2, Singapore}%
\affiliation{MajuLab, CNRS-UNS-NUS-NTU International Joint Research Unit, UMI 3654, 117543, Singapore}
\affiliation{Department of Electrical and Computer Engineering, 
 National University of Singapore}%

\date{\today}

\begin{abstract}
  We investigate work extraction protocols designed to transfer the maximum possible energy to a battery using sequential access to $N$ copies of an unknown pure qubit state. The core challenge is designing interactions to optimally balance two competing goals: charging of the battery optimally using the qubit in hand, and acquiring more information by qubit to improve energy harvesting in subsequent rounds. Here, we leverage
   exploration-exploitation trade-off in reinforcement learning to develop adaptive strategies achieving energy dissipation that scales only poly-logarithmically in $N$. This represents an exponential improvement over  current protocols based on full state tomography. 
\end{abstract}

\maketitle
\end{CJK*}

\noindent\textbf{Introduction}\m
Given sequential access to finite, identical samples of an unknown quantum system, what is the optimal strategy for extracting work from them and charging a battery? 
The extraction of useful work from available resources has long been a central problem in classical thermodynamics and continues to attract significant attention in the quantum domain. While numerous protocols have been proposed~ \cite{allahverdyan2004maximal,aaberg2013truly,brandao2013resource,skrzypczyk2014work,elouard2018efficient}, the majority assume that the agent has full knowledge of the quantum state---these are so-called \emph{state-aware} protocols. In practice we may not always know how the state is prepared, leading us to the \emph{state-agnostic} scenario at hand. In this case, how can do we design a protocol for work extraction?

A natural strategy might involve first performing quantum state tomography to estimate the unknown state, followed by work extraction based on this estimate \cite{vsafranek2023work,watanabe2024black,watanabe2025universal}. Indeed, knowledge of the quantum state is essential for efficient work extraction; consider, for example, Szilard's engine, where the information about the system's configuration---such as the position of a particle---enables work to be extracted \cite{szilard1929entropieverminderung}. However, with only finitely many samples available, any estimate of the true state has statistical uncertainty~\cite{o2016efficient,haah2016sample}, resulting in unavoidable heat dissipation during work extraction \cite{riechers2021initial}. Furthermore, quantum systems that are measured during state tomography are no longer available for work extraction and therefore contribute to irreversible heat dissipation. Combining these observations, we will show that any two-step procedure that first uses tomography to inform an efficient work extraction procedure will lead to a cumulative dissipation that scales at least as $\Omega(\sqrt{N})$ in the number $N$ of copies of the unknown state.
Such strategies are only minimally adaptive though and this raises a natural question: \emph{Can more adaptive strategies, which simultaneously balance learning and work extraction, offer better performance in terms of cumulative dissipation?} We answer this in the affirmative, showing an exponential improvement for pure qubit states.

\begin{figure}[htbp!]
    \centering
    \includegraphics[width=0.9\linewidth]{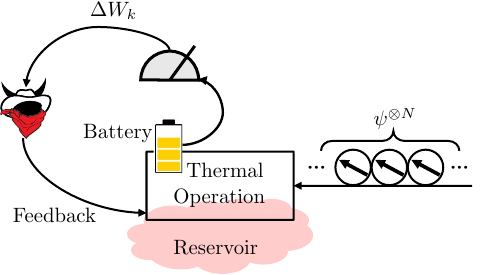}
    \caption{ Sketch of the sequential work extraction protocol with a thermal reservoir. At each time step $k \in [N]$, the agent receives a copy of an unknown qubit state $\psi$ and performs a thermal operation involving the reservoir and a battery. A measurement in the battery system is carried out to determine the extracted work $\Delta W_k$, which is then used as feedback to improve the extraction strategy in subsequent rounds.
}
    \label{fig:protocol}
\end{figure}

More specifically, we consider a source that sequentially emits a total of $N$ identical unknown pure qubit states, which are used to charge a battery using an interaction that couples the qubit and battery to a thermal reservoir. For this we adopt a semi-classical thermodynamic model where the battery is represented as a classical weight that stores energy via vertical displacement. We then design a \emph{state-agnostic} protocols that controls the interaction between the qubit and the battery depending on data collected in previous rounds and examine how the cumulative dissipation scales with the number of available samples. We find that adaptive strategies can result in an exponential improvement of the cumulative dissipation: whereas tomography-based algorithms yield a cumulative dissipation that scales as $\Omega(\sqrt{N})$, we show that a scaling of $O(\polylog(N))$ is possible for suitable adaptive work-extraction algorithms for pure states. Our main technical contribution is an upper bound on the cumulative dissipation in this model. This result builds on a novel connection between adaptive quantum control and classical reinforcement learning~\cite{lattimore2020bandit}, which we believe to be of independent interest. More precisely, we show that the dissipation in our framework can be interpreted as the \emph{regret} incurred during the learning of pure quantum states, formulated within a multi-armed bandit setting~\cite{lumbreras22bandit,lumbreras24pure}.

\noindent\textbf{Extracting work from knowledge}\m
We consider a commonly studied extraction model, where the goal is to extract energy from a qubit system in contact with a thermal bath~\cite{skrzypczyk2014work,huang2023engines}. Energy extraction is quantified by its use to raise the potential energy of an external battery --- represented by the lifting of an external weight. Here, we assume that the qubit system is initially unknown. The three physical subsystems involved in the process are:

\begin{itemize}[leftmargin=*]
    \item Unknown pure state source (System $A$): A qubit in a pure state $\psi_A = \ketbra{\psi}{\psi}$ with a degenerate Hamiltonian $H_A = \omega \id/2$. This is the system from which free energy is to be extracted.
    
    \item Battery (System $B$): A semi-classical weight described by a continuous-variable state $\varphi(x) \in L^2(\mathbb{R})$. The battery Hamiltonian is defined as $H_B \varphi(x) = x \varphi(x)$, where $x$ represents the height of the weight. The energy of the battery can be changed by translating the weight up by a certain height $x_0$, described by the translation operator $\Gamma^B_{x_0} \varphi(x) = \varphi(x-x_0)$.
    
    \item Thermal reservoir (System $R$): A heat bath at a fixed inverse temperature $\beta$, modeled as a supply of qubit states $\gamma_\beta(\nu) = Z_R(\nu)^{-1} e^{-\beta H_R(\nu)}$, where the Hamiltonian is $H_R(\nu) = \nu \ketbra{1}{1}$ and $Z_R(\nu) = \tr(e^{-\beta H_R(\nu)})$ is the partition function. Here, $\ket{0}$ and $\ket{1}$ are energy eigenstates, and $\nu$ is the energy gap. This energy gap can be tuned.
\end{itemize}
We consider a degenerate system Hamiltonian, such that every orthogonal basis is an energy eigenbasis.  This choice avoids the phenomenon of ``work-locking" --- the inability of thermal operations to extract work from coherence in the energy eigenbasis~\cite{lostaglio2015description}. In contrast, systems with non-degenerate Hamiltonian requires additional energy expenditure. Further details are provided in Appendix~\ref{apd:first_model}.

We consider an agent with oracle access to the unknown system $A$ over a finite number of rounds $N\in\mathbb{N}$, indexed by $k \in [N]$. The goal is to convert the maximum amount of free energy from $A$ and store it in the battery $B$ through interactions with a thermal reservoir $R$. However, since the state $\psi$ of the system is unknown, the agent cannot extract work optimally from the outset. Instead, it must gradually improve its strategy by learning from each round, where the learning is done by measuring the battery. The work extraction protocol is defined by two key components: a policy that updates the agent’s guess $\psi_k$ of the unknown state $\psi$ at each round, and a sequence $\{ \epsilon_k \}_{k=1}^N$ that determines the accuracy of these guesses. Both quantities can be chosen adaptively and together they determine the dissipation at each round. We first present the general structure of the protocol and later discuss how the dissipation depends on these choices, and how to optimize them for best performance. The protocol proceeds as follows at each round $k \in [N]$:\\

\setlist{nolistsep}
\begin{enumerate}[itemsep=4pt,leftmargin=*] 
    \item The agent receives a sample of the unknown qubit state $\psi$.
    
    \item Based on the outcomes from previous rounds, the agent selects a direction $\psi_k$ on the Bloch sphere, sets an accuracy $\epsilon_k\in [0,1]$ and defines a basis $\lbrace \psi_k , \psi_k^\perp \rbrace$ for system $A$. This computation is done using the previously selected directions and measured battery energies $\lbrace \psi_s , \mu_s \rbrace_{s=1}^{k-1}$.
    \item We first implement a unitary on the system qubit in the form of 
    \begin{equation}
        U_k = \ketbra{0}{\psi_k} + \ketbra{1}{\psi_k^\perp},
    \end{equation}
    satisfying $[H_A,U_k]=0$ which tries to diagonalize the system qubit in computational basis.
    \item 
    The agent then performs a thermal operation by repeatedly appending a reservoir qubit $R$, applying an energy-conserving unitary on the combined system $A B R$ to transfer energy from $R$ to the battery $B$ and discarding $R$. Specifically, for each of $M$ repetitions (indexed by $\tau \in [M]$), the agent:
    \begin{itemize}[leftmargin=*]
        \item Sets the energy gap of the reservoir qubit to $\nu (\tau,\epsilon_k )$ (see Eq.~\eqref{eq:intro_gap_parametrization}) and gets a fresh qubit state $\gamma_\beta(\nu(\tau,\epsilon_k))$.
        \item Applies the following unitary 
        \begin{align}\label{eq:unitary_intro}
            V_{\psi_k , \tau} = \sum_{i,j} \ket{i}\!\bra{j}_A \otimes \ket{j}\!\bra{i}_R \otimes \Gamma^B_{(i-j)\nu(\tau,\epsilon_k)} ,
        \end{align}
       satisfying $[H_A+H_B+H_R,V_{\psi_k,\tau}]=0$ to $ABR$.
       
       \item Discards the reservoir qubit.
    \end{itemize}
    
    \item After completing the $M$ steps, the agent measures the energy of the battery in its eigenbasis and records the energy $\mu_k$.
\end{enumerate}

\begin{figure}
    \centering
    \includegraphics[width=0.9\linewidth]{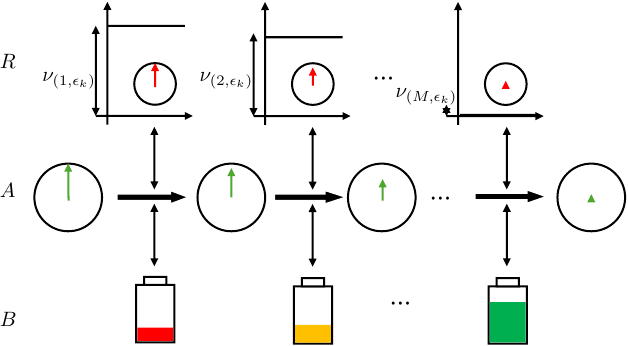}
    \caption{Illustration of the repetitions of the thermal operation in the full system $ABR$, where arrows represent Bloch vectors of states, showing that the system qubit becomes more and more mixed as the process goes. The energy gaps $\{\nu_{k,i}\}_i$ of successive reservoir Hamiltonian forms a strictly decreasing sequence, making the successive thermal states more mixed. At each step, we take a new qubit from the reservoir and swap the system qubit with the reservoir qubit, the energy from reservoir will flow into the battery.
    At the end of the process, the qubit in system $A$ is the thermal state.}
    \label{fig:protocol_pic}
\end{figure}

For the above protocol, the expected work extracted at round $k \in [N]$, as a function of the chosen direction $\psi_k$ and accuracy parameter $\epsilon_k$, is given by
\begin{align}
\label{eq:avgwork}
 \Ex[\Delta W_k] = \beta^{-1} \left[ \D (\psi \| \id/2) - \D (\psi \| \Delta_{2\epsilon_k}(\psi_k)) \right],
\end{align}
where $\Delta_{\epsilon}(\rho) = (1 - \epsilon)\rho + \epsilon \id/2$ denotes the depolarizing channel and $\D(\rho\|\sigma) = \tr(\rho(\ln \rho- \ln\sigma ))$ is the quantum relative entropy. The maximal expected work per round, $\beta^{-1} \D(\psi \| \mathbb{I}/2)$, is achieved when the agent perfectly estimates the state, i.e., when $\psi_k = \psi$ and $\epsilon_k = 0$. Accordingly, we define the dissipation at round $k$ as:
\begin{align}
\dissipation^{k} := \max_{\psi_k, \epsilon_k} \mathbb{E}[\Delta W_k] - \mathbb{E}[\Delta W_k] = \beta^{-1} \D(\psi \| \Delta_{2\epsilon_k}(\psi_k)).
\end{align}
This expression highlights a key trade-off: to reduce dissipation, the agent must align $\psi_k$ with the true state $\psi$, but it cannot set $\epsilon_k$ too small unless the estimate is sufficiently accurate, as otherwise the divergence becomes large. The parameter $\epsilon_k$ thus plays a dual role, quantifying both the uncertainty in the estimate and its thermodynamic penalty. The agent's objective is to extract the maximum amount of work into the battery using the $N$ copies of the unknown system. Equivalently, the agent aims to minimize the cumulative dissipation over $N$ rounds, which is given by
\begin{align}\label{eq:dissipation_sc}
    \dissipation(N) = \beta^{-1} \sum_{k=1}^N \dissipation^{k}=\sum_{k=1}^N  \D(\psi\| \Delta_{2\epsilon_k} (\psi_k)) .
\end{align}

Going back to the protocol we note that the unitary $V_{\psi_k,\tau}$ in Eq.~\eqref{eq:unitary_intro} swaps the states of the unknown system $A$ and the thermal reservoir $R$. This operation induces an energy exchange determined by the energy gap $\nu (\tau,\epsilon_k )$ between the two systems. The energy difference is transferred to the battery through a translation operation $\Gamma^B_{(i-j)\nu (\tau,\epsilon_k )}$ to conserve the total energy of the combined system.

To maximize the extractable non-equilibrium free energy from system $A$, it is essential to minimize dissipation during the energy transfer to the battery. This can be achieved by ensuring that the process is quasi-static --- that is, slow and nearly reversible. Under such conditions, the system qubit remains close to the thermal state of the reservoir's Hamiltonian, thereby suppressing heat flow during the interaction.  Conversely, if the process is carried out rapidly, i.e., the system state deviates from reservoir's thermal state, the system will appear out of thermal equilibrium with the reservoir, resulting in heat exchange, which contributes to the entropy production during the protocol. The fundamental trade-off between efficiency and power in such protocols is well established: in general, the reduction in work output scales as $1/\tau_0$ where $\tau_0$ denotes time taken for extraction protocol~\cite{van2022finite}. This can also be characterized by discrete number of interactions required to implement the transformation~\cite{taranto2023landauer}.

To approximate a quasi-static process, we repeat the thermal interaction $M$ times within each round. In the limit $M \to \infty$, the process becomes effectively reversible. At each repetition $\tau \in [M]$, we introduce a slight mismatch between the energy levels of system $A$ and the reservoir $R$ by varying the energy gap. We use the following parametrization of the gap
\begin{align}\label{eq:intro_gap_parametrization}
    \nu (\tau,\epsilon_k ) = \beta^{-1} 
    \ln \left( \frac{1 - \frac{\tau}{2M}- \left(1 - \frac{\tau}{M} \right) \epsilon_k }{\frac{\tau}{2M}+\left(1 - \frac{\tau}{M} \right) \epsilon_k} \right),
\end{align}
where $\epsilon_k \in [0,1]$ controls the mismatch. The choice of $\epsilon_k$ will be discussed later for specific protocols. Further details on this gap parametrization can be found in Appendix~\ref{apd:first_model}. The full interaction of systems $A$, $B$, and $R$ through the unitary $V_{\psi_k,\tau}$ is illustrated in Figure~\ref{fig:protocol_pic}.

To compute the total extracted work, we evaluate at each repetition $\tau\in[M]$ how much energy is transferred to the battery. This is done by considering the two initial input states that can trigger a swap operation, namely $\ket{0}_A\ket{1}_R$ and $\ket{1}_A\ket{0}_R$. When the unitary Eq.~\eqref{eq:unitary_intro} is applied repeatedly, there is a certain probability that the states of the system and reservoir qubits are swapped, thereby enabling energy exchange with the battery. In the quasi-static limit $M \rightarrow \infty$, each of these initial states contributes a fixed amount of energy $\Delta W_k = \mu_{k+1} - \mu_k$ to the battery, which are respectively
\begin{align}
    \label{eq:work_values_intro}
    \begin{split}
        w_{k,0} &:=  \beta^{-1}(\D(\psi_{k}\|\id/2) +  \ln (1-\epsilon_k)), \\
        w_{k,1} &:= \beta^{-1}(\D(\psi_{k}^\perp\|\id/2) +  \ln \epsilon_k).
    \end{split}
\end{align}

To build intuition for $w_{k,i}$, consider $w_{k,0}$ as an example. This represents the extracted work given $\psi_k$. The relative entropy $\D(\psi_k\|\id/2)$ reflects the non-equilibrium free energy of $\psi_k$, which corresponds to the extracted work in the quasi-static limit. The logarithmic term arises because the process is not entirely quasi-static --- it is quasi-static for all but the first repetition. During the first repetition, the system qubit $\ketbra{0}{0}$ is swapped with the reservoir qubit $(1-\epsilon_k)\ketbra{0}{0} + \epsilon_k \ketbra{1}{1}$. This swap generates entropy $\Delta S=-\epsilon_k\ln \epsilon_k-(1-\epsilon_k)\ln(1-\epsilon_k)$ and yields work extraction $\Delta U =-\beta^{-1}\epsilon_k\nu(1,\epsilon_k) =-\beta^{-1}\epsilon_k\ln\frac{\epsilon_k}{1-\epsilon_k}$. Combining both, the resulting change in free energy of the system and the battery is $\Delta U- \beta^{-1}\Delta S = \ln (1-\epsilon_k)$, which precisely accounts for the logarithmic term. A similar argument holds for $w_{k,1}$.

We now consider the case where the initial state is $\psi$, the total energy extracted will be distributed as
\begin{align}\label{eq:work_distribution_intro}
  \mathrm{Pr}(  \Delta W_k = w_{k,i} ) = 
  i+ (-1)^i|\langle \psi | \psi_k \rangle |^2  ,
\end{align}
which combined with Eq.~\eqref{eq:work_values_intro} gives the expected extracted work stated in Eq.~\eqref{eq:avgwork}. More details of this computation are provided in Appendix~\ref{apd:first_model}.

\noindent \textbf{Extracting work after learning}\m
An attentive reader may notice that achieving sublinear dissipation in the number of copies \( N \) is, in principle, not difficult. The sequence of battery measurement outcomes follows the distribution Eq.~\eqref{eq:work_distribution_intro}, which corresponds to a measurement in the direction $\psi_k$ on the unknown state $\psi$. Hence, these outcomes can be used to perform state tomography of $\psi$ and then select $\psi_k$ and $\epsilon_k$ that minimizes $D(\psi \| \Delta_{2\epsilon_k} (\psi_k))$. A natural strategy is to allocate a fraction $\alpha N$, with \( 0 \leq \alpha \leq 1 \), of the available copies to learning the unknown state and producing an estimate that will fix a direction $\hat{\psi}$ and $\hat{\epsilon}$, then use the remaining $(1-\alpha)N$ copies for work extraction by setting $\psi_k = \hat{\psi}$ and $\epsilon_k = \hat{\epsilon}$. However, such a two-phase approach is fundamentally limited by the sample complexity needed to achieve accuracy in relative entropy. Standard tomography algorithms are not designed to minimize dissipation during learning and typically result in linear dissipation. The cumulative dissipation for this approach can be lower bounded as
\begin{align}
\dissipation(N) = \Omega\left( \alpha N + (1- \alpha)N\epsilon_{\alpha N} \right),
\end{align}
where \( \epsilon_{\alpha N} \) is the smallest achievable relative entropy error using \( \alpha N \) copies. This error is lower bounded as \( \epsilon_{\alpha N} = \Omega(\frac{1}{\alpha N}) \) for $\alpha N$ samples~\cite{haah2016sample,Flammia2024quantumchisquared,chen_adaptivity}, which implies $\dissipation(N) = \Omega\left(\alpha N + \frac{1}{\alpha}\right)$. Minimizing this expression over \( \alpha \) yields a lower bound of \( \dissipation(N) = \Omega(\sqrt{N}) \). Moreover, a matching upper bound \( \dissipation(N) = O(\sqrt{N}) \) can be achieved using an optimal learning algorithm for relative entropy~\cite{Flammia2024quantumchisquared}. This naturally raises the question of whether the dissipation for pure states can be further reduced through more adaptive or integrated strategies, or whether the $\Omega(\sqrt{N})$ scaling already represents the fundamental limit in this setting.

\noindent \textbf{Extracting work while learning}\m Indeed, using a more adaptive approach, it is possible to surpass the $\sqrt{N}$ barrier. To achieve this, we use the fully adaptive state tomography algorithm studied in~\cite{lumbreras24pure}, which is based on the multi-armed bandit algorithm from~\cite{pmlr-v247-lumbreras24a}. The algorithm assumes sequential access to $N$ identical copies of an unknown pure qubit state $|\psi\rangle$ and performs adaptive single-copy measurements using rank-$1$, two-outcome POVMs. At each round $k\in[N]$, the algorithm selects a direction $\psi_k$ and performs a projective measurement in the basis $\lbrace \psi_k, \psi_k^\perp \rbrace$. The binary outcome after the measurement or ``reward'' $r_k\in\lbrace 0,1\rbrace$ is distributed according to Born's rule
\begin{align}
    \label{eq:reward_intro_bandit}
    \mathrm{Pr} (R_k = r_k) = 
    \begin{cases}
    |\langle\psi_k|\psi\rangle|^2, \quad  & r_k=1,\\
    1-|\langle\psi_k|\psi\rangle|^2, \quad & r_k=0.
    \end{cases}
\end{align}
These outcomes match the energy distribution of the thermal protocol Eq.~\eqref{eq:work_distribution_intro}, and can be used within that framework by setting $r_k = i$ if $\Delta W_k = w_{k,1-i}$ for $i\in\lbrace 0,1 \rbrace$.

The directions of the measurements are chosen to ensure low infidelity with the unknown state, which can be converted into a bound on the relative entropy. This can be done by selecting $\epsilon_k$ such that $1 - |\langle \psi | \psi_k \rangle|^2 \leq \epsilon_k \leq \frac{1}{2}$ with high probability and then bound $\D(\psi\|\Delta_{2\epsilon_k} (\psi_k))$. This can be done applying~\cite[Proposition 2.34]{Flammia2024quantumchisquared} which leads to
\begin{align}\label{eq:dissipaiton_infidelity}
       \dissipation(N) \leq \beta^{-1}\sum_{k=1}^N 16\epsilon_k (2-\ln \epsilon_k).
\end{align}

Now we review the exact updating rule  of $\psi_k$ at each time step $k\in\{1,2,\ldots,N\}$. Each $\psi_k$ is adaptively based on the previous rewards $\lbrace r_s \rbrace_{s=1}^{k-1}$ and measured directions $\lbrace \psi_s \rbrace_{s=1}^{k-1}$. The updating rule is formulated in the Bloch vector representation. The algorithm works in $K^*\in\mathbb{N}$ stages where for each stage $k^*\in[K^*]$ performs $t$ measurements on the Bloch vector directions  $\lbrace a_{k^*,i}  \rbrace_{i=1}^4 \subset \mathbb{R}^3$ which we are going to define. The total number of rounds is then $N=4tK^*$.  The sampled reward are then used to build the following $t$ weighted least-square estimators
\begin{align}
\widetilde{\theta}_{k^*,j}^\text{w} = V_{k^*}^{-1}  \sum_{l = 1}^{k^*} \frac{1}{\hat{\sigma}_l^2} \sum_{i=1}^{4} a_{l,i}(2r_{l,i,j}-1) , 
\end{align}
where $r_{l,i,j}\in\lbrace 0,1\rbrace$ is the reward for the measurement direction along $a_{l,i}$ for $l\in[K^*]$ and $j\in[t]$. Then $\hat{\sigma}_l^2$ is an upper bound on the variance of $r_{l,i,j}$ and $V_{k^*}$ is the design matrix 
\begin{align}
V_{k^*} = \id+\sum_{l = 1}^{k^*}\frac{1}{\hat{\sigma}_{l}^2} \sum_{i=1}^{4} a_{l,i}  (a_{l,i})^\intercal .
\end{align}
As studied in~\cite{pmlr-v247-lumbreras24a,lumbreras24pure}, a weighted least-squares estimator boosts learning along directions with low variance, which are precisely those close to $\psi$, since $\mathrm{Var}(r_k) \leq 1 - |\langle \psi_k | \psi \rangle |^2$. For the specific update used by the algorithm, it can be shown that $\hat{\sigma}_{l}^2 \sim 1/\sqrt{\lambda_{\max}(V_{k^*-1})}$, (where $\lambda_{\max}(\cdot)$ denotes the maximum eigenvalue) provides an upper bound on the outcome variances and ensures well-defined least-squares estimators. However, a single estimator is not sufficient to guarantee proximity to the unknown state. For this reason, the algorithm constructs multiple estimators and combines them into a normalized weighted median-of-means $\theta^{\text{wMoM}}_{k^*}$, which satisfies the required probabilistic guarantees to be close to $\psi$. Details on the construction of $\theta^{\text{wMoM}}_{k^*}$ are provided in Appendix~\ref{apd:pure_state_regret}. Finally it outputs the measurement directions on the Bloch sphere representation  $\{a_{k^*+1,i}\}_{i=1,2,3,4}$ of
\begin{align}\label{eq:measuremts_update}
     \tilde{a}_{k^*+1,i} = {\theta}^{\text{wMoM}}_{k^*} - \frac{(-1)^{i}}{\sqrt{\lambda_{\min}(V_{k^*})}}v_{k^*,\lceil\frac{i}{2}\rceil},
\end{align}
where $\lambda_{\min}(V_{k^*})$ is the minimum eigenvalue of $V_{k^*}$ and $\lbrace v_{k^*,0}, v_{k^*,1} \rbrace$ are the two eigenvectors of $V_{k^*}$ with smallest eigenvalues. The normalized Bloch vectors $a_{k^*+1,i}$ are used to update the measurement directions $\psi_k$ for round $k$ in stage $k^*+1$. More details can be found in Appendix~\ref{apd:pure_state_regret}.

It has been proven~\cite[Theorems 9 and 11]{lumbreras24pure} that, with high probability, the algorithm achieves an infidelity scaling of $1 - |\langle\psi_k| \psi\rangle|^2 \leq C \ln(N) / k$ for all $k \in \{1,2,\ldots,N\}$. The same bound also holds in expectation. We can thus use this guarantee to set $\epsilon_k = C \ln(N) / k$ in the thermal protocol, using the measurement directions $\psi_k$ given by Eq.~\eqref{eq:measuremts_update}, and since $\epsilon_k$ upper bounds the infidelity between $\psi$ and $\psi_k$ we can use the dissipation bound Eq.~\eqref{eq:dissipaiton_infidelity} to obtain the following result.

\noindent \textbf{Main Result}\m
There exists an explicit protocol for the semi-classical battery model that adaptively updates the direction $\psi_k$ and the accuracy $\epsilon_k$ based on the rewards $\lbrace r_s \rbrace_{s=1}^{k-1}$, achieving, with probability at least $1-\delta$ 
\begin{align} 
\dissipation (N) = O\left(\beta^{-1}\ln^2(N) \ln \left( \frac{N}{\delta} \right)  \right) .
\end{align} 

Notice that in this model, measurements in the energy eigenbasis of the battery were performed. According to Laudauer's principle, such measurements reduce entropy and thus incur an energetic cost associated with information erasure. This form of dissipation --- arising from cost of resetting memory---should also be taken into account. In Appendix~\ref{sec:landauer} we show that the dissipation required by Landauer's principle also scales as $O(\polylog(N))$ when our adaptive algorithm is employed. 

Furthermore, this algorithm is not specific to the work extraction protocol discussed above. We have also applied it to an alternative protocol modeled using the Jaynes-Cummings Hamiltonians, where the battery is represented as a uniform energy ladder. In this setting, the work dissipation similarly exhibits  $O(\polylog(N))$ scaling. Interested readers are referred to Appendix~\ref{apd:jc_protocol} for further details.

\noindent\textbf{Conclusion and outlook}\m
We are the first to frame sequential work extraction as an instance of the exploration-exploitation trade-off and to quantify cumulative dissipation in the finite-copy regime. Beyond establishing a provable upper bound on the total dissipation, we explicitly construct an adaptive algorithm that achieves it. Our approach not only advances the understanding of work extraction under partial information but also opens a pathway for extending this framework to the extraction of other quantum resources beyond free energy—such as entanglement, coherence, and magic—within their respective resource theories.

\textbf{Acknowledgments} JL and RCH contributed equally to this work. JL thanks Jan Seyfried for discussions about quantum state tomography with quantum relative entropy loss.  This work is supported by the National Research Foundation of Singapore through the NRF Investigatorship Program (Award No. NRF-NRFI09-0010), the National Quantum Office, hosted in A*STAR, under its Centre for Quantum Technologies Funding Initiative (S24Q2d0009), grant FQXi-RFP-IPW-1903 (``Are quantum agents more energetically efficient at making predictions?") from the Foundational Questions Institute (FQxI) and Fetzer Franklin Fund (a donor-advised fund of Silicon Valley Community Foundation),  the Singapore Ministry of Education Tier 1 Grant RT4/23, MT, YH and JL are supported by the National Research Foundation, Singapore and A*STAR under its CQT Bridging Grant and its Quantum Engineering Programme (NRF2021-QEP2-02-P05)

\bibliographystyle{ultimate}
\bibliography{biblio_qbandits_qthermo}

\newpage

\onecolumngrid
\pagenumbering{arabic}

\appendix

\section{Learning pure quantum states (almost) without regret}\label{apd:pure_state_regret}

In this Section we review the quantum state tomography algorithm presented in~\cite{lumbreras24pure}. They considered sequential access to an unknown pure quantum state $\ket{\psi}$ and at each time step $k\in\{1,\ldots,N\}$, a reward measurement is performed on the direction $\ket{\psi_k}$. Formally the reward measurement is described by a rank-1 two-outcome POVM $\lbrace \psi_k, \psi_k^\perp \rbrace$ where $\psi_k=\ketbra{\psi_k}{\psi_k}$ corresponds to $R_k=1$ and $\psi_k^\perp=\mathbb{I}-\psi_k$ corresponds to $R_k=0$. The observed reward $R_k$ is distributed accordingly to Born's rule, i.e 
\begin{align}\label{eq:prob_quantum_reward}
    \mathrm{Pr} \left(R_k = r_k \right) = 
    \begin{cases}
    |\langle \psi_k | \psi \rangle |^2, \quad & r_k=1,\\
    1-|\langle \psi_k | \psi \rangle |^2, \quad & r_k=0. 
    \end{cases}
\end{align}
The goal of this work was to design an algorithm that uses measurements that minimally disturb the unknown state $\ket{\psi}$. They used as a figure of merit for this task the regret which is defined as
\begin{align}\label{eq:infidelity_regret}
    \regret (N) = \sum_{k=1}^N (1 - |\langle \psi_k | \psi \rangle|^2),
\end{align}
and is the cumulative sum of infidelities between the unknown state $|\psi\rangle$ and the selected direction $\psi_k$. The algorithm is formulated in terms of the Bloch vector which means that at each time step $k$, the algorithm uses the previous and current information of the reward $\lbrace r_1,\ldots , r_{k} \rbrace$ and corresponding directions $\lbrace \ket{\psi_1},\ldots , \ket{\psi_{k}} \rbrace $ to output a vector $a_{k+1}\in\mathbb{R}^3$ that is normalized $\| a_{k+1} \| =1$ and it is linked to the next reward measurement described by the two-outcome POVM  $\lbrace \psi_{k+1} , \psi_{k+1}^\perp \rbrace$ (or the direction $\ket{\psi_{k+1}}$) as
\begin{align}
    \psi_k = \frac{1}{2} (\mathbb{I} + a_k \cdot \sigma ),
\end{align}
where $\sigma = (\sigma_x , \sigma_y , \sigma_z )$ are the Pauli matrices (or any other basis of $2\times 2$ Hermitian matrices). The pseudo-code for how the algorithm updates the reward measurements can be found in Algorithm~\ref{alg:linucb_vn_var} (\textsf{LinUCB Vanishing Variance Noise}) and now we describe all the involved quantities. The algorithm presented in~\cite{lumbreras24pure} was formulated for a more general case but here we focus for the specific qubit case. 

\begin{algorithm}[H]
	\caption{\textsf{LinUCB-VVN}} 
	\label{alg:linucb_vn_var}
 
        Require: $\lambda_0\in\mathbb{R}_{>0}$, $t\in\mathbb{N}$
        
        Set initial design matrix $V_0 \gets \lambda_0\mathbb{I}$. 
        
        Set initial estimate of variance $\hat{\sigma}^2_{1} \gets 1$. 
        
        Set initial Bloch vectors $a_{1,1}=(\frac{1}{\sqrt{2}},0,\frac{1}{\sqrt{2}})^\intercal$, $a_{1,2}=(-\frac{1}{\sqrt{2}},0,\frac{1}{\sqrt{2}})^\intercal$, $a_{1,3}=(0,\frac{1}{\sqrt{2}},\frac{1}{\sqrt{2}})^\intercal$, $a_{1,4}=(0,-\frac{1}{\sqrt{2}},\frac{1}{\sqrt{2}})^\intercal$.

        \For{$k^*=1,2,\ldots$}{
            \vspace{1mm}
            \textit{Optimistic action selection}
            \vspace{1mm}
            
            \For{$i = 1,2,3,4$}{               
                \vspace{1mm}
                \textit{Perform $t$ independent measurements for each $a_{k^*,i}$} 
                \vspace{1mm}
                
                \For{$j=1,...,t$}{
                    Measure the unknown $|\psi\rangle$ in the Bloch vector directions given by $a_{k^*,i}$ and receive outcomes $r_{k^*,i,j}$ 
                }
            }

            \vspace{1mm}
            \textit{Update design matrix}
            \vspace{1mm}
            
            $V_{k^*} \gets V_{k^*-1} + \frac{1}{\hat{\sigma}_{k^*}^2} \sum_{i=1}^{4} a_{k^*,i}  a_{k^*,i}^\intercal $        
            
            \vspace{1mm}
            \textit{Update LSE for each subsample}
            \vspace{1mm}

            \For{$j=1,2,...,t$:}{
                $\widetilde{\theta}_{k^*,j}^\text{w} \gets V_{k^*}^{-1}  \sum_{s = 1}^{k^*} \frac{1}{\hat{\sigma}_{k^*}^2} \sum_{i=1}^{4} a_{s,i} (2r_{s,i,j}-1)$
            } 
            
            \vspace{1mm}
            \textit{Update Bloch vectors for estimates}
            \vspace{1mm}
            
            Compute ${\theta}_{k^*}^{\text{\tiny wMOM}}$ using $\lbrace \widetilde{\theta}_{k^*,j}^\text{w} \rbrace_{j=1,\ldots t}$ according to Eq.~\eqref{eq:median_of_means}

            \vspace{1mm}
            \textit{Update Bloch vectors for measurements}
            \vspace{1mm}
            
            Select Bloch vectors $a_{k^*+1,i}$ using ${\theta}_{k^*}^{\text{\tiny wMOM}}$ according to Eq.~\eqref{eq:action_general_update}

            \vspace{1mm}
            \textit{Update estimator of variance for $a_{k^*+1,i}$}
            \vspace{1mm}
            
            $\hat{\sigma}^2_{k^*+1} \gets \frac{2\zeta}{\sqrt{\lambda_{\max}(V_{k^*})}}$
        }
\end{algorithm}

The algorithm described in Algorithm~\ref{alg:linucb_vn_var} takes as input a parameter $t$ that is used to control the success probability and updates the measurements in stages. Each stage $k^*$ contains $4$ groups labeled by $i\in\{1,2,3,4\}$, and each group $i$ contains $t$ rounds labeled by $j\in\{1,\ldots,t\}$. Each round $k$ can also be labeled by its stage $k^*$, group $i$ and group index $j$ where $k=4t(k^*-1)+4(j-1)+i$. At each time stage $k^*$, the algorithm takes $4$ Bloch vectors $\lbrace a_{k^*,i} \rbrace_{i=1,\ldots,4}$ and performs $t$ independent reward measurements defined by $a_{k^*,i}$ in each group $i$. The updates for $a_{k^*,i}$ can be defined recursively given the previous information of outcomes and measurements. At each stage $k^*$ the algorithm computes the following $t$ weighted least-square estimators of the Bloch vector of the unknown state $\ket{\psi}$ as
\begin{align}
\widetilde{\theta}_{k^*,j}^\text{w} = V_{k^*}^{-1}  \sum_{l = 1}^{k^*} \frac{1}{\hat{\sigma}_l^2} \sum_{i=1}^4a_{l,i}( 2r_{l,i,j}-1),
\end{align}
where $r_{l,i,j}\in\lbrace 0 ,1 \rbrace$ is the outcome of the reward measurement defined by $a_{l,i}$ in round $(l,i,j)$, $\hat{\sigma}_{l}^2$ is the estimate of reward variances in stage $l$ and $V_{k^*}$ is the design matrix defined as
\begin{align}
  V_{k^*} = V_{k^*-1} + \frac{1}{\hat{\sigma}_{k^*}^2} \sum_{i=1}^{4} a_{k^*,i}  a_{k^*,i}^\intercal,
\end{align}
where the initial design matrix is setup to be $V_0 = \lambda_0 \mathbb{I}$ for some parameter $\lambda_0 > 0$. Then the algorithm does an extra step that outputs a normalized version ${\theta}^{\text{wMoM}}_{k^*}$ of median of means with the $k$ weighted least squares estimator $\widetilde{\theta}^{\text{wMoM}}_{k^*}$ in the following way 
\begin{align}
    \label{eq:median_of_means}
    {\theta}^{\text{wMoM}}_{k^*} = \frac{\widetilde{\theta}^{\text{wMoM}}_{k^*}}{\| \widetilde{\theta}^{\text{wMoM}}_{k^*} \|_2} 
\end{align}
where
\begin{align}
        \widetilde{\theta}_{k^*}^{\text{\tiny wMoM}} := \tilde{\theta}_{k^*,j^*}, 
\end{align}
where     
\begin{align}
    j^* = \argmin_{j\in[t]} \text{median}\lbrace \|\tilde{\theta}_{t^*,j} - \tilde{\theta}_{t^*,j'} \|_{V_{k^*}}: j'\in [k]/j \rbrace,
\end{align}
and $\| x \|^2_{V_{k^*}} = \langle x, V_{k^*} x\rangle $ is the weighted norm given by $V_{k^*}$ and $\langle \cdot , \cdot \rangle$ is the standard inner product between real vectors. Then the update of the measurements is defined as
\begin{align}\label{eq:action_general_update}
    a_{k^*+1,i} = \frac{\widetilde{a}_{k^*+1,i}}{\| \widetilde{a}_{k^*+1,i}\|_2}, 
\end{align}
where 
\begin{align}
    \tilde{a}_{k^*+1,i} = {\theta}^{\text{wMoM}}_{k^*} - \frac{(-1)^i}{\sqrt{\lambda_{\min}(V_{k^*})}}v_{k^*,\lceil\frac{i}{2}\rceil}
\end{align}
where $\lbrace v_{k^*,0},v_{k^*,1}\rbrace$ are the two eigenvectors of $V_{k^*}$ with smallest eigenvalues, $\lambda_{\min}(V_{k^*})$ is the smallest eigenvalue of $V_{k^*}$ and the estimate of reward variances $\hat{\sigma}^2_{k^*}$ is set to
\begin{align}\label{eq:weight_choice}
    \hat{\sigma}^2_{k^*} = \frac{2\zeta}{\sqrt{\lambda_{\max}(V_{k^*})}},
\end{align}
where $\lambda_{\max}(V_{k^*})$ is the largest eigenvalue of $V_{k^*}$ and $\zeta$ is any constant such that $\zeta \geq 334812\sqrt{2}+1296\sqrt{6}$. The particular form $\hat{\sigma}^2_{k^*}$ given by Eq.~\eqref{eq:weight_choice} guarantees that $\hat{\sigma}^2_{k^*}$ is a good upper bound for the variance of the outcomes $r_{k^*,i,j}$ that are sampled after performing a measurement on the directions given by $a_{k^*,i}$. With this choice as argued in~\cite{lumbreras24pure} it allows to obtain rigorous concentration bounds such that $\widetilde{\theta}^{\text{wMoM}}_{k^*} $ is a good estimator for the Bloch vector of the unknown state $\ket{\psi}$.

Now we state the main theorem from~\cite{lumbreras24pure} that states how the regret scales with the time horizon $N$ and also how the infidelities between the unknown state $\ket{\psi}$ the directions $|\psi_k\rangle$ scale with respect to the unknown state $\ket{\psi}$.

\begin{theorem}[{\cite[Theorem 9 and 11]{lumbreras24pure}}]\label{th:fidelity_gurantee_bandit}
 Fix $K^*\in\mathbb{N}$,  $t= \lceil 24\ln \left( K^* / \delta \right) \rceil $ for some $\delta > 0$ and time horizon $N = 4t K^*$. Then we have that the quantum state tomography Algorithm~\ref{alg:linucb_vn_var} over an unknown state $\ket{\psi}$ achieves with probability at least $1-\delta$ the regret Eq.~\eqref{eq:infidelity_regret} scaling
\begin{align}
    \regret (N) \leq C_1 \ln \left( \frac{N}{\delta} \right)\ln ( N ),
\end{align}
 for some universal constant $C_1 > 0$. Also for all $k\in\{1,\ldots,N\}$ the selected 2-outcome POVM's given by the rank-1 projector $\psi_k = \ketbra{\psi_k}{\psi_k}$ achieve infidelity
 \begin{align}
   1- |\langle \psi_k | \psi \rangle|^2 \leq C_2 \frac{\ln\left(\frac{N}{\delta} \right)}{k},
 \end{align}
 for some universal constant $C_2 > 0$. Moreover setting $\delta = \frac{1}{N^*}$ it holds
 \begin{align}
     \Ex [\regret (N)] = C_3\ln^2 (N), \quad \Ex [1- |\langle \psi_k | \psi \rangle|^2] \leq C_4 \frac{\ln (N) }{k},
 \end{align}
 for some universal constants $C_3 , C_4 > 0$ and the expectation is taken over the probability distribution of outcomes and measurements induced by the policy. 
\end{theorem}

\section{\texorpdfstring{$\rho^*$}{}-work extraction protocol}
\label{apd:first_model}

Here we discuss more details about the work-extraction protocol that is used, we adapted the protocol formalized in Skrzypczyk's paper~\cite{skrzypczyk2014work}. Just as in their formulation, the expected work extracted from a known state $\rho$ will precisely be given by the state's non-equilibrium free energy, which equals the relative entropy between the state and Gibbs' state, $\gamma_\beta$ with inverse temperature $\beta$, i.e.
\begin{equation}
    \label{eq:relative_entropy}
    \beta\Ex(W) = \D(\rho\|\gamma_\beta). 
\end{equation}
We will be applying the protocol to a degenerate Hamiltonian. Discussion on non-degenerate Hamiltonian will be discussed towards the end of the section.

We focus on a specific time step within the $N$ rounds of extraction, doing so simplifies notation by removing the $k$ index. In a specific round, the agent is given a partially unknown qubit system state $\psi$, and a classical description of the direction $\hat{\psi}$ and an accuracy $\epsilon$. The agent will then choose to optimize the protocol for a state $\rho^* = (1-\epsilon)\hat{\psi}+ \epsilon\hat{\psi}^\perp$.
Along the unknown state $\psi$, he also has access to a heat bath at inverse temperature $\beta$ and a battery state $\varphi(x)$. The Hamiltonian of the system is $H_A = \omega \id/2$. The heat bath can provide any amount of thermal state with any Hamiltonian at inverse temperature $\beta$. We will mainly consider qubit thermal states $\gamma_\beta(\nu) = \frac{1}{Z_R(\nu)}e^{-\beta H_R(\nu)}$ where $Z_R(\nu) = \tr(e^{-\beta H_R(\nu)})$ with Hamiltonian $H_R(\nu)=\nu\ketbra{1}{1}$ where $\{\ket{i}\}_{i=0,1}$ are the energy eigenstates. The battery is modeled as a weight at a certain height whose state is described by $\varphi(x)\in L^2(\mathbb{R})$ and Hamiltonian $H_B$ such that $H_B \varphi(x-x_0) = x\varphi(x-x_0)$. We will assume $\varphi(x-x_0)$ to be a battery state whose energy is sharply centered at $x_0$. The energy of the battery can be changed by translating the weight up by a certain height $x_0$, described by the translation operator $\Gamma_{x_0}^B \varphi(x) = \varphi(x-x_0)$. We aim to extract work from the partially unknown system system state into the battery. We will design the work extraction protocol for the state $\rho^*$, i.e. the protocol optimally extracts work from $\rho^*$, see Algorithm~\ref{alg:sc_work_extraction}. 

 To simplify calculation as well as maintain generality later, we will denote $\psi$ as $\rho$, $\hat{\psi}$ as $\ket{\phi_0}$ and $\hat{\psi}^\perp$ as $\ket{\phi_1}$, likewise we denote $p_0 =1-\epsilon$ and $p_1=\epsilon$, so $\rho^* = \sum_i p_i\ketbra{\phi_i}{\phi_i}$.

\begin{algorithm}[H]
	\caption{\textsf{$\rho^*$-ideal work extraction}} 
	\label{alg:sc_work_extraction}
 
        Require: An unknown system state $\psi$, a classical description of the state $\rho^*$, a battery state $\varphi(x)$ with the battery energy $\mu=0$, a reservoir at inverse temperature $\beta$

        Set $\{\phi_{i}\}_i$ and $\{p_{i}\}_i$ to be the eigenvectors and eigenvalues of $\rho^*$

        \vspace{1mm}
        \textit{Unitary Rotation}
        \vspace{1mm}
        
        Apply unitary $U=\sum_i \ketbra{i}{\phi_i}$ to try to diagonalize the system qubit in computational basis.
        \vspace{1mm}
        
        \For{$\tau=1,2,\ldots,M$}{
            \vspace{1mm}
            \textit{Prepare a fresh a reservoir qubit and exchange it with the system}
            \vspace{1mm}
            
            Take a fresh thermal qubit $\gamma_\beta(\nu(\tau,\epsilon))=\frac{1}{Z_R(\nu(\tau,\epsilon))}e^{-\beta H_R(\nu(\tau,\epsilon))}$ where $\nu(\tau,\epsilon)=\beta^{-1}\ln\frac{p_{0,\tau}}{p_{1,\tau}}$, $p_{i,\tau} = p_i-(-1)^i \tau \delta p$ and $\delta p=\frac{1}{M}(p_0-\frac{1}{2})$ from the reservoir.  
            
            Apply the swap unitary $V_{\rho^*,\tau}= \sum_{ij}\ketbra{i}{j}_A\otimes\ketbra{j}{i}_R \otimes \Gamma_{(i-j)\nu(\tau,\epsilon)}$ on the system, the battery and the reservoir qubit. 
            
            Discard the reservoir qubit. 
        }
        
        \vspace{1mm}
        \textit{Measure the extracted work}
        \vspace{1mm}
            
        Measure the battery energy, obtain the battery energy $\mu'$ and compute the extracted work $\Delta W=\mu'-\mu$. 
\end{algorithm}

In the first stage of the protocol, we rotate the unknown qubit via unitary 
\begin{equation}
    U = \sum_i\ketbra{i}{\phi_i}~.
\end{equation}
This operation attempts to diagonalize the system qubit in the computation basis.
We then interact the system with the battery. The state of the system together with the battery is 
\begin{equation}
\label{eq:intial_joint_state}
    \rho_{AB} = \sum_{ij}\bra{\phi_i}\rho\ket{\phi_j}\ketbra{i}{j}_A\otimes \varphi(x)_B.
\end{equation}

In the second stage of the protocol, we perform $M$ repetitions of the following process. In repetition $\tau$, we take a fresh thermal qubit $\gamma_\beta(\nu(\tau,\epsilon))$ with Hamiltonian $H_{R}(\nu(\tau,\epsilon))$ from the reservoir where $\nu(\tau,\epsilon)= \beta^{-1} \ln \frac{p_{0,\tau}}{p_{1,\tau}}$, $p_{i,\tau}= p_{i} - (-1)^i \tau \delta p$ and $\delta p = \frac{1}{M}(p_{0} - \frac{1}{2})$. Note that the reservoir qubit we take depends on which repetition we are in. Then we perform the swap unitary 
\begin{align}
    V_{\rho^*,\tau}  = \sum_{ij}\ketbra{i}{j}_A\otimes\ketbra{j}{i}_R \otimes \Gamma_{(i-j)\nu(\tau,\epsilon)}~.
\end{align}
This unitary swaps the system and the fresh qubit from the reservoir, extracts work into the battery due to the different energy gap between $\{\ket{i}_A\}_{i=0,1}$ and $\{\ket{i}_R\}_{i=0,1}$ and conserves energy of the system, the qubit from the reservoir and the battery. Finally, the qubit from the reservoir is discarded. At the end of each repetition $\tau$, the reduced state is 
\begin{align}
    \rho_{AB,\tau}  = \tr_R\left(V_{\rho^*,\tau}\left(\rho_{AB,\tau-1}\otimes \gamma_\beta(\nu(\tau,\epsilon))\right) V_{\rho^*,\tau}^\dagger\right),
\end{align}
After the first repetition, we obtain 
\begin{align}
    \label{eq:joint_state_1}
    \rho_{AB,1} = \sum_{i} p_{i,1} \ketbra{i}{i}_A \otimes \rho_{B,i,1},
\end{align}
where 
\begin{align}
    \label{eq:battery_state_1}
    \rho_{B,i,1} = \sum_j \bra{\phi_{j}}\rho\ket{\phi_{j}} \varphi(x-(i-j)\nu(\tau,\epsilon)), 
\end{align}
and after repetition $\tau$ where $\tau\geq 2$, we obtain
\begin{align}
    \label{eq:joint_state_tau}
    \rho_{AB,\tau} = \sum_{i} p_{i,\tau} \ketbra{i}{i}_A \otimes \rho_{B,i,\tau}, 
\end{align}
where
\begin{align}
    \label{eq:battery_state_tau}
    \rho_{B,i,\tau} = \sum_j  p_{j,\tau-1} \Gamma_{(i-j)\nu(\tau,\epsilon)} \rho_{B,j,\tau-1} \Gamma_{(i-j)\nu(\tau,\epsilon)}^\dagger. 
\end{align}
From Eq.~\eqref{eq:joint_state_tau}, we observe that the reduced state of the system changes gradually, which resembles a quasi-static process in thermodynamics. This is the reason why we take the swap unitary in repetition.

\subsection{Work distribution}
\label{apd:work_distribution}
In this section, we will use the Lagrange mean value theorem and the first mean value theorem for definite integrals~\cite{strang2019calculus} as follows:
\begin{theorem}
    \label{thm:lag_mean_value_theorem}
    Let $f:[a, b] \to \mathbb{R}$ be a continuous on the closed interval $[a,b]$ and differentiable on the open interval $(a,b)$. Then there exists $c\in (a, b)$ such that
    \begin{align}
        f(b)-f(a) = f'(c) (b-a). 
    \end{align}
\end{theorem}
\begin{theorem}
    \label{thm:int_mean_value_theorem}
    Let $f:[a, b] \to \mathbb{R}$ be a continuous function on the closed interval $[a,b]$. Then there exists $c\in (a, b)$ such that
    \begin{align}
        \int_a^b f(x) \textnormal{d} x = f(c) (b-a). 
    \end{align}
\end{theorem}

We will show the following theorem in the following subsection.
\begin{theorem}
    \label{thm:work_distribution}
    Let $\lbrace\phi_i\rbrace_{i=0,1}$ and $\lbrace p_i\rbrace_{i=0,1}$ be the eigenvectors and eigenvalues of $\rho^*$, $\Delta W$ be the extracted work (which is a continuous random variable) and $M$ be the number of repetitions as in Algorithm~\ref{alg:sc_work_extraction}. It holds that: if the extraction protocol is operated on a state $\rho$ that is the eigenstate of $\rho^*$, i.e., $\rho=\phi_{i}$, then then the expected extracted work $\Ex[\Delta W]$ converges to a fixed value $w_i$ and the extracted work $\Delta W$ converges in probability to its expectation $\Ex[\Delta W]$. To be precise, it means
        \begin{align}
            \lim_{M\to \infty} \Ex[\Delta W] = w_i,
        \end{align}
        where
        \begin{equation}
        \label{eq:thm3work_values}
            w_{i} \coloneqq \beta^{-1} (\D(\phi_{i}\|\id/2) + \ln p_{i}),
        \end{equation} 
        and for any $\epsilon>0$
        \begin{align}
            \lim_{M\to\infty} \Pr[|\Delta W - \Ex[\Delta W]|\geq \epsilon ] = 0.
        \end{align}

\end{theorem}
\begin{proof}
We consider the case where $\rho=\phi_{i}$. We assume that $p_{0} > \frac{1}{2}$ as we deal with an estimate for a pure state in Algorithm~\ref{alg:sc_work_extraction}, although similar proof holds for other cases. According to Eq.~\eqref{eq:joint_state_1}, the state after the first repetition can be viewed as a classical state described as follows: the state after repetition $1$ is $\phi_{x_1}$ where $x_1$ is a random bit sampled from $\{0,1\}$ according to the probability distribution $(p_{0,1},p_{1,1})$; the extracted work after the first repetition conditioned on $x_1$ is $(x_1-i)\nu(1,\epsilon)$. According to Eq.~\eqref{eq:joint_state_tau}, the evolution in repetition $\tau$ where $\tau\geq 2$ can be viewed as a classical process described as follows: the state after repetition $\tau$ is $\phi_{x_\tau}$ where $x_{\tau}$ is a random bit sampled from $\{0,1\}$ according to the probability distribution $(p_{0,\tau}, p_{1,\tau})$; the extracted work in repetition $\tau$ conditioned on $x_{\tau-1}x_\tau$ and is $(x_\tau - x_{\tau-1}) \nu(\tau,\epsilon)$. Suppose that the random bits sampled during the above process is $x_1\ldots x_M$ after $M$ repetitions. The extracted work after repetition $M$ conditioned on $x_1\ldots x_M$ is
\begin{align}
\label{eq:B9}
    \Delta W= (x_1-i)\nu(1,\epsilon) + \sum_{\tau=2}^M (x_\tau - x_{\tau-1})\nu(\tau,\epsilon) = - i  \nu(1,\epsilon) + \sum_{\tau=1}^{M-1} x_\tau (\nu(\tau,\epsilon) - \nu(\tau+1,\epsilon)) + x_M \nu(M,\epsilon), 
\end{align}
recall that 
\begin{align}
\label{eq:nu_tau}
    \nu(\tau,\epsilon) = \beta^{-1} \ln \frac{p_{0} - \tau \delta p }{p_{1} + \tau \delta p} . 
\end{align}
where $\delta p = \frac{1}{M}(p_{0}-\frac{1}{2})$.
The expected extracted work is 
\begin{align}
    \Ex[\Delta W] =   - i  \nu(1,\epsilon) + \sum_{\tau=1}^{M-1} \Ex[x_\tau] (\nu(\tau,\epsilon) - \nu(\tau+1,\epsilon)) + \Ex[x_M] \nu(M,\epsilon).
\end{align}
Notice that, from the definition of $x_\tau$, $\Ex[x_\tau] = p_{1,\tau}=p_{1} + \tau \delta p$, 
we obtain
\begin{align}
\label{eq:B12}
    \Ex[\Delta W] & =   - i \beta^{-1}\ln \frac{p_{0} - \delta p }{p_{1} + \delta p} + \beta^{-1} \sum_{\tau=1}^{M-1}  \left(  \ln \frac{p_{0} - \tau \delta p }{p_{1} + \tau \delta p} - \ln \frac{p_{0} - (\tau+1) \delta p }{p_{1} + (\tau+1) \delta p}\right) (p_{1}+\tau\delta p ). 
\end{align}
We now use the definition of $\nu(\tau,\epsilon)$ in Eq.~\eqref{eq:nu_tau} as well as the Lagrange mean value theorem in Theorem~\ref{thm:lag_mean_value_theorem} to obtain 
\begin{align}
    \beta (\nu(\tau,\epsilon) - \nu(\tau+1,\epsilon)) = \ln \frac{p_{0} - \tau \delta p }{p_{1} + \tau \delta p} -\ln \frac{p_{0} - (\tau+1) \delta p }{p_{1} + (\tau+1) \delta p} = \frac{1}{\xi_\tau(1-\xi_\tau)}  \delta p, 
\end{align}
for some $\xi_\tau\in [p_{0}-(\tau+1) \delta p, p_{0}-\tau\delta p ]$. 

Therefore, Eq.~\eqref{eq:B12} can be simplified to 
\begin{equation}
\begin{split}
    \label{eq:expected_extracted_work}
    \Ex[\Delta W]  & =  - i  \beta^{-1} \left(\ln \frac{p_{0} }{p_{1}} -\frac{\delta p}{\xi_0 (1-\xi_0)}  \right)+ \beta^{-1} \sum_{\tau=1}^{M-1}   \frac{p_{1}+\tau\delta p }{\xi_\tau(1-\xi_\tau)}  \delta p   \\
    & =  - i  \beta^{-1} \ln \frac{p_{0} }{p_{1}} + \beta^{-1} \sum_{\tau=1}^{M} \frac{p_{1}+\tau\delta p }{\xi_\tau(1-\xi_\tau)}  \delta p \\
    & \quad + i  \beta^{-1} \frac{\delta p}{\xi_0 (1-\xi_0)}  - \beta^{-1} \frac{\delta p}{2\xi_M (1-\xi_M)}~.
\end{split}
\end{equation}

We will approximate the sum in the second line of Eq.~\eqref{eq:expected_extracted_work} with an integration, where the remainder is bounded due to first mean value theorem for definite integrals as in Theorem~\ref{thm:int_mean_value_theorem}. Namely, 
\begin{align}
    \beta^{-1} \sum_{\tau=1}^{M}  \frac{p_{1}+\tau\delta p }{\xi_\tau(1-\xi_\tau)}  \delta p =\beta^{-1} \int_{\frac{1}{2}}^{p_{0} } \frac{\text{d} p}{p} + \beta^{-1}R_1(\delta p) = \beta^{-1}\ln p_{0}+ \beta^{-1}\ln(2) + \beta^{-1}R_1(\delta p),
\end{align}
where $R_1(\delta p)$ is the remainder given by 
\begin{align}
    R_1(\delta p) & = \sum_{\tau=1}^{M}  \frac{p_{1}+\tau\delta p }{\xi_\tau(1-\xi_\tau)}  \delta p -  \int_{\frac{1}{2}}^{p_{0}} \frac{\text{d}p}{p} = \sum_{\tau = 1}^{M} \left( \frac{p_{1}+\tau\delta p }{\xi_\tau(1-\xi_\tau)}  \delta p - \int_{p_{0}-\tau \delta p}^{p_{0}-(\tau-1)\delta p}\frac{\text{d} p}{p} \right) \\ 
    & = \sum_{\tau=1}^{M} \left(\frac{p_{1}+\tau\delta p }{\xi_\tau(1-\xi_\tau)} -  \frac{1}{\xi_\tau'} \right)\delta p =  \sum_{\tau=1}^{M} \frac{\xi_\tau(1-\xi_\tau) - \xi_\tau'(p_{1}+\tau\delta p)}{\xi_\tau'\xi_\tau(1-\xi_\tau)} \delta p, 
\end{align}
where from the first line to the second line, we have used the first mean value theorem for definite integrals~Theorem~\ref{thm:int_mean_value_theorem} that
\begin{align}
    \int_{p_{0}-\tau \delta p}^{p_{0}-(\tau-1)\delta p}\frac{\text{d} p}{p} = \frac{1}{\xi_\tau'} \delta p,
\end{align}
for some $\xi_\tau'\in [p_{0}-\tau \delta p, p_{0}-(\tau-1)\delta]$. Therefore, the remainder satisfies
\begin{align}
    |R_1(\delta p )| & \leq \sum_{\tau=1}^{M} \left|\frac{\xi_\tau(1-\xi_\tau) - \xi_\tau'(p_{1}+\tau\delta p)}{\xi_\tau'\xi_\tau(1-\xi_\tau)} \right|\delta p \leq \sum_{\tau=1}^{M} \left|\frac{p_{1}+\tau\delta p + \xi_\tau}{\xi_\tau'\xi_\tau(1-\xi_\tau)} \right| (\delta p)^2  \leq \sum_{\tau=1}^{M}\frac{4}{p_{1}} (\delta p)^2 \leq  (2p_{0}-1)\frac{2}{p_{1}}\delta p. 
\end{align}

Therefore, the second line in Eq.~\eqref{eq:expected_extracted_work} is finite while the third line is infinitesimal, and we obtain
\begin{align}
    \Ex[\Delta W] & =  - i  \beta^{-1} \ln \frac{p_{0} }{p_{1}} + \beta^{-1}\ln p_{0}+ \beta^{-1}\ln(2) + O(\delta p) \\
    & = -\beta^{-1} \tr(\phi_{i}\ln\id/2)  - i  \beta^{-1} \ln \frac{p_{0} }{p_{1}} + \beta^{-1}\ln p_{0} + O(\delta p) \\
    & = -\beta^{-1} \tr(\phi_{i}\ln\id/2) + \beta^{-1}\ln p_{i} + O(\delta p) \\
    & = \beta^{-1}\left[\D\left(\phi_{i}\middle\|\id/2\right) + \ln p_{i} \right] + O(\delta p). 
\end{align}
Since $\delta p \propto\frac{1}{M}$, we then obtain that
\begin{align}
    \Ex[\Delta W] =  \beta^{-1}\left[\D\left(\phi_{i}\middle\|\id/2\right) + \ln p_{i} \right] + O\left(\frac{1}{M}\right).
\end{align}
Taking $M\to \infty$, we obtain
\begin{align}
    \lim_{M\to \infty}\Ex[\Delta W] =  \beta^{-1}\left[\D\left(\phi_{i}\middle\|\id/2\right) + \ln p_{i} \right].
\end{align}
Now we demonstrate the convergence of $\Delta W$ towards its expectation value, recall from Eq.~\eqref{eq:B9}, we have that 
\begin{align}
    \Delta W =  - i  \nu(1,\epsilon) + \sum_{\tau=1}^{M-1} x_\tau (\nu(\tau,\epsilon) - \nu(\tau+1,\epsilon)) + x_M \nu(M,\epsilon).  
\end{align}
By the Lagrange mean value theorem,  
\begin{align}
     \nu(\tau,\epsilon) - \nu(\tau+1,\epsilon) = \ln \frac{p_{0} - \tau \delta p }{p_{1} + \tau \delta p} -\ln \frac{p_{0} - (\tau+1) \delta p }{p_{1} + (\tau+1) \delta p} = \frac{\delta p}{\xi_\tau(1-\xi_\tau)} , 
\end{align}
for some $\xi_\tau \in [p_{0}-(\tau+1)\delta p ,p_{0}-\tau\delta p ] $ and $\tau=1,\ldots, (M-1)$ satisfying 
\begin{align}
    |\nu(\tau,\epsilon) - \nu(\tau+1,\epsilon)| \leq \frac{2}{p_{1}} \delta p.
\end{align}
We thus obtain that $x_\tau(\nu(\tau,\epsilon) - \nu(\tau+1,\epsilon))\in [0,\frac{2}{p_{1}} \delta p]$ for $\tau=1,\ldots, (M-1)$. Besides, $\nu(M,\epsilon)=0$. The convergence rate to the expectation, by the Hoeffding inequality, is given by 
\begin{align}
    \Pr[|\Delta W - \Ex[\Delta W]|\geq \varepsilon ] \leq 2 e^{-\frac{\varepsilon^2}{\sum_{\tau=1}^{M-1} \left(\frac{2}{p_{1}} \delta p \right)^2}} \leq 2 e^{-\frac{p_{1}^2 \varepsilon^2 M}{(2p_{0}-1)^2}}. 
\end{align}
Taking $M\to \infty$, we obtain
\begin{align}
    \lim_{M\to\infty}\Pr[|\Delta W - \Ex[\Delta W]|\geq \epsilon ] =0. 
\end{align}
\end{proof}
Theorem~\ref{thm:work_distribution} demonstrates that the extracted work is close to either $w_{0}$ or $w_{1}$,  
with probability close to $\bra{\phi_{0}}\rho\ket{\phi_{0}}$ and $\bra{\phi_{1}}\rho\ket{\phi_{1}}$ respectively. Therefore, measuring the extracted work $\Delta W$ from the state $\rho$ in Algorithm~\ref{alg:sc_work_extraction} is effectively measuring the state $\rho$ in the basis $\{\phi_{i}\}_{i=0,1}$ up to an error probability exponentially vanishing with respect to the number of repetitions $M$ in Algorithm~\ref{alg:sc_work_extraction}. When $\rho$ is a pure state i.e., $\rho=\ketbra{\psi}{\psi}$, the energy measurement of the battery is equivalent to the reward measurement in the quasi-static limit of Algorithm~\ref{alg:linucb_vn_var}, and their correspondence is (without loss of generality assuming $w_{0}\geq w_{1}$)
\begin{align}
    r = \begin{cases}
        1,\quad & \Delta W \geq \frac{w_{0}+w_{1}}{2}, \\
        0,\quad & \Delta W \leq \frac{w_{0}+w_{1}}{2}. 
    \end{cases}
\end{align}
The distribution of the reward is 
\begin{align}
    \Pr[R=r] = \begin{cases}
        |\langle \phi_0|\psi\rangle|^2+\epsilon_{\text{error}},\quad & r=1, \\
        1- |\langle\phi_{0}|\psi\rangle|^2 - \epsilon_{\text{error}},\quad & r=0, 
    \end{cases}
\end{align}
where $|\epsilon_{\text{error}}| \leq O(e^{-CM})$ for some constant $C$. In the limit of $M\to\infty$, the correspondence reduces to 
\begin{align}\label{eq:reward_thermal}
    r = \begin{cases}
        1,\quad & \Delta W = w_{0}, \\
        0,\quad & \Delta W = w_{1}, 
    \end{cases}
\end{align}
and the distribution of the reward reduces to 
\begin{align}
   \Pr[R=r] = \begin{cases}
        |\langle \phi_0|\psi\rangle|^2,\quad & r=1, \\
        1- |\langle\phi_{0}|\psi\rangle|^2 ,\quad & r=0, 
    \end{cases}
\end{align}
Note that as Algorithm~\ref{alg:sc_work_extraction} proceeds with increasing $k$, the error probability that the algorithm can tolerate in the dreward measurement decreases polynomially, which requires the number of repetitions $M$ increases poly-logarithmically.

\subsection{Extracted work for different inputs}
\begin{theorem}\label{th:work_differnt_input}
\label{thm:6} 
Let $\lbrace\phi_i\rbrace_{i=0,1}$ and $\lbrace p_i\rbrace_{i=0,1}$ be the eigenvectors and eigenvalues of $\rho^*$ and $w_i$ be the value of work extracted defined in Theorem.~\ref{thm:work_distribution}. It holds that,
\begin{enumerate}
    \item When applying the protocol to any state $\rho$, the probability of measuring $\Delta W = w_i$ 
        is given by 
        \begin{equation}
        \label{eq:thm3work_probs}
            \Pr(\Delta W=w_i)=\bra{\phi_{i}}\rho\ket{\phi_i}~.
        \end{equation}

    \item When the extraction protocol is operated on a state $\rho$ where $\rho\neq \rho^*$, the expected work extracted is given by
        \begin{equation}
            \Ex[\Delta W]=\beta^{-1}\left[\D(\rho\middle\|\id/2)-\D(\rho\|\rho^*)\right]~,
        \end{equation}
        where the second term can be defined as the the dissipation due to the agent's imperfect knowledge of $\rho$.
\end{enumerate}
\end{theorem}
\begin{proof}
We first observe that the off-diagonal term $\bra{\phi_i}\rho\ket{\phi_j} \ketbra{i}{j}_A\otimes \varphi(x)_B$ of the join state in Eq.~\eqref{eq:intial_joint_state} does not affect Eq.~\eqref{eq:joint_state_1}. Therefore, it is identical for the case where the input is $\rho$ and the case where the input is $\sum_i \bra{\phi_i}\rho\ket{\phi_i} \phi_i$. The latter case can be viewed as a probabilistic mixture of cases where the input is $\phi_{i}$ with probability $\bra{\phi_{i}}\rho\ket{\phi_{i}}$. Therefore, Statement 1 in Theorem~\ref{thm:6} holds, i.e.,
\begin{equation}
    \Pr(\Delta W=w_i)=\bra{\phi_{i}}\rho\ket{\phi_i}~.
\end{equation}

Next, to prove Statement 2, we consider the case where a protocol optimized for $\rho^*=\sum_ip_i\ketbra{\phi_i}{\phi_i}$ is applied onto an arbitrary state $\rho$ with possibly $\rho\neq\rho^*$. Using Eq.~\eqref{eq:thm3work_values} and~\eqref{eq:thm3work_probs}, the extracted work is given by 
\begin{align}
    \Ex[\Delta W] & = \bra{\phi_{0}} \rho \ket{\phi_{0}} \beta^{-1} \left[\D\left(\phi_{0}\|\id/2\right) + \ln p_{0}\right] + \bra{\phi_{1}} \rho \ket{\phi_{1}} \beta^{-1} \left[\D\left(\phi_{1}\|\id/2\right) + \ln p_{1}\right] \\
    & = \beta^{-1}\left[\bra{\phi_{0}} \rho \ket{\phi_{0}}  (-  \tr(\phi_{0}\ln\id/2)+  \ln p_{0})- \bra{\phi_{1}} \rho \ket{\phi_{1}}   (-\tr(\phi_{1}\ln \id/2)+  \ln p_{1} )\right] \\
    & = \beta^{-1}\left[\tr(\mathcal{P}(\rho)\ln\rho^*) -\tr(\mathcal{P}(\rho)\ln \id/2) \right] = \beta^{-1} \left[ \tr(\rho\ln\rho^*) - \tr(\rho\ln\id/2) \right],
\end{align}
where $\mathcal{P}(\rho) = \sum_i \phi_{i} \rho \phi_{i}$ the pinching map.
We can also express the expected work in term of relative entropy:
\begin{align}
\label{eq:b42}
    \Ex[\Delta W]  =  \beta^{-1}\left[ \tr(\rho\ln\rho^*) - \tr(\rho \ln\id/2)\right] = \beta^{-1}\left[\D\left(\rho\|\id/2\right)-\D(\rho\|\rho^*)\right]~.
\end{align}
\end{proof}

Note that in the event $\rho^*=\rho$, i.e., the agent is fully aware of the identity of the quantum state and is able to ensure the work extraction protocol to be entirely quasi-static. Then the extract work is given by  
\begin{align}
    \Ex[\Delta W] & = p_{0} \beta^{-1} \left[\D\left(\phi_{0}\middle\|\id/2\right) + \ln p_{0}\right] + p_{1} \beta^{-1} \left[\D\left(\phi_{1}\|\id/2\right) + \ln p_{1}\right] \\
    & = \beta^{-1}(- p_{0} \tr(\phi_{0}\ln\id/2)+ p_{0}\ln p_{0} -   p_{1}  \tr(\phi_{1}\ln\id/2)+ p_{1}\ln p_{1} ) \\
    & = \beta^{-1}\left[ \tr(\rho \ln \rho) - \tr(\rho\ln\id/2)\right] = \beta^{-1} \D\left(\rho\|\id/2\right). 
\end{align}
We retrieve the full non-equilibrium free energy.
Therefore, the dissipation due to agent's imperfect knowledge of the true state $\rho$ can be quantified as 
\begin{align}
    \dissipation = \max_{\rho^*}\Ex[ \Delta W] - \Ex[ \Delta W] = \beta^{-1} \D(\rho\|\rho^*). 
\end{align}

\subsection{Cumulative dissipation}

In this section, we consider the setting where we have oracle sequential access to an unknown pure qubit state $\psi$, and our goal is to extract the maximal amount of work into a battery system. To achieve this, we can use Algorithm~\ref{alg:linucb_vn_var} with the rewards~\eqref{eq:reward_thermal} to learn an approximate direction of the state, and then run Algorithm~\ref{alg:sc_work_extraction} to extract work based on the approximate input. In general, we can consider mixed-state estimator $\hat{\rho}_k$. Assuming sequential access to the unknown state over $N$ rounds, and using the expected extracted work from Theorem~\ref{th:work_differnt_input}, we define the dissipation at round $k \in [N]$ with respect to the optimal protocol as
\begin{align}
\dissipation^{k} := \beta^{-1} \D(\psi \| \hat{\rho}_k ),
\end{align}
and the cumulative dissipation over all $N$ rounds as
\begin{align}\label{eq:apendix_dissipation_sc}
\dissipation(N) := \beta^{-1} \sum_{k=1}^N \dissipation^{k} = \beta^{-1} \sum_{k=1}^N \D(\psi \| \hat{\rho}_k).
\end{align}

\begin{algorithm}[H]
	\caption{\textsf{Thermal work extraction}}
	\label{alg:cum_dissipation}
        Require: sequence $\lbrace \epsilon_k \rbrace_{k=1}^\infty$
          
	\For {$k=1,2,\ldots$}{
            Receive unknown $ | \psi \rangle$ and couple to battery state $\varphi(x-\mu_{k-1})$
           
            Compute direction $|\psi_k \rangle$ with Algorithm~\ref{alg:linucb_vn_var} using $\lbrace \psi_s , r_s \rbrace_{s=1}^{k-1}$

            Set $\hat{\rho}_k = \Delta_{2\epsilon_k} ( \psi_k )$
           
            Extract work using Algorithm~\ref{alg:sc_work_extraction} with input $\hat{\rho}_k$ and get extracted work $\Delta W_k$ and energy $\mu_k$
           
            Set reward $r_k = \lbrace 0 , 1 \rbrace$ according to~\eqref{eq:reward_thermal}
        }
\end{algorithm}

To minimize the cumulative dissipation, we use Algorithm~\ref{alg:cum_dissipation}, which takes as input a sequence of accuracies $\lbrace \epsilon_k \rbrace_{k=1}^\infty$. At each round, the estimator uses $\psi_k$, the direction output by Algorithm~\ref{alg:linucb_vn_var}, and sets $\hat{\rho}_k = \Delta_{2\epsilon_k}(\psi_k)$, where $\Delta_{2\epsilon_k}$ is the completely depolarizing channel. If $\epsilon_k$ is a good approximation of the infidelity between the true state $\psi$ and the estimate $\psi_k$, then the dissipation $\dissipation^k$ is controlled by $\epsilon_k$. This is formalized in the following theorem.
\begin{theorem}[Theorem 2.34 in~\cite{Flammia2024quantumchisquared}]\label{th:relative_entropy_fideity} 
Let $\rho$ and $\hat{\rho}$ be $d$-dimensional quantum states achieving infidelity $1 - F(\rho,\hat{\rho}) \leq \epsilon \leq \frac{1}{2}$. Then we have
\begin{align}
    \D ( \rho \| \Delta_{2\epsilon} ( \hat{\rho}) ) \leq 16\epsilon \left( 2 + \ln \left( \frac{d}{2\epsilon} \right) \right) . 
\end{align}
\end{theorem}

Given the above bound we can use the fidelity guarantee of Algorithm~\ref{alg:linucb_vn_var} in Theorem~\ref{th:fidelity_gurantee_bandit} to prove a bound on the cumulative dissipation.

\begin{theorem} Given a finite time horizon $N\in\mathbb{N}$ and $\delta\in (0,1 )$ there exists an explicit sequence of accuracies $\lbrace \epsilon_k \rbrace_{k=1}^\infty $ such that Algorithm~\ref{alg:cum_dissipation} achieves
\begin{align} 
\dissipation (N) = O\left(\beta^{-1}\ln^2(N) \ln \left( \frac{N}{\delta} \right)  \right) .
\end{align} 
\end{theorem}

\begin{proof}
    We can choose
    \begin{align}
        \epsilon_k = \min \left\lbrace C \frac{\ln\left(\frac{N}{\delta} \right)}{k} ,\frac{1}{2} \right\rbrace ,
    \end{align}
    where $C$ is the constant in Theorem~\ref{th:fidelity_gurantee_bandit} for the fidelity bound of the direction $\psi_k$. Then we can use Theorem~\ref{th:relative_entropy_fideity} combined with the fidelity guarantee of Theorem~\ref{th:fidelity_gurantee_bandit} to get that with  probability at least $1-\delta$ we have
    \begin{align}
        \dissipation (N) \leq \beta^{-1} \sum_{k=1}^{N} 16 \epsilon_k (2-\ln \epsilon_k ).
    \end{align}
The result follows by noting that, for a sufficiently large constant $k^*$, we have $\epsilon_k = C \frac{\ln\left(\frac{N}{\delta} \right)}{k}$ for all $k \geq k^*$. The dissipation incurred during the first $k^*$ rounds contributes a constant term. For the remaining rounds $k \geq k^*$, using $-\ln\epsilon_k\leq \ln N$ for $k\leq N$ and summing the corresponding dissipation terms yields the claimed polylogarithmic scaling.
\end{proof}

\subsection{Non-degenerate Hamiltonian}
So far we have discussed the case for when the system Hamiltonian is degenerate, in general though we can consider Hamiltonian's with energy gap of $\omega$, i.e., $H_A = \omega\ket{E_1}\!\bra{E_1}$. In this case, there is a pre-defined energy eigenbasis, hence all the swapping operations will have to be done in such basis. As mentioned before, under the constraint of strict energy conservation, there is no way for one to extract the full non-equilibrium free energy if the initial state is not diagonalized in the energy eigenbasis. However, it is possible if we relax the constraint to energy conservation on average, in which case a unitary rotation, $U_{\rho^*}$ can be carried out on the system qubit before the swap operations were applied. The energy difference inccured during the rotation can be shifted into the battery via the unitary, 
\begin{equation}
    U_{\rho^*} = \sum_i \ketbra{E_i}{\phi_{i}} \otimes \Gamma_{e_i}^B~,
\end{equation}
where $e_i = \tr(\phi_{i}H_A) -\omega_i$ is the energy difference.
This operation obeys average energy conservation as long as the input is diagonalized along the eigenbasis of $\{\phi_{i}\}_i$. However, when the input is not diagonalized, which is usually the case where input is unknown, this operation is non-energy conserving even on the average and hence will require additional energy be supplied. The amount of additional work required is at least the energetic difference between the unknown state and the pinched version of it. Although a possible experimental setup was proposed in \cite{skrzypczyk2014work}, it is based on the implementation of a time-dependent interaction Hamiltonian. This in turn would require additional resources to keep track of time and change the Hamiltonian smoothly \cite{woods2023autonomous}. More recent work has shown that such an operation can only be realized if one has access to unbounded coherence \cite{aaberg2014catalytic,korzekwa2016extraction}, while this can potentially be achieved approximately using lasers, the energy from the laser also has to be accounted for.  In order to avoid including unnecessary technicality and shift the focus of the paper, we focus on the degenerate Hamiltonian.

\section{Jaynes-Cummings work extraction protocol}
\label{apd:jc_protocol}

\begin{algorithm}[H]
        \caption{Jaynes-Cummings work extraction} 
	\label{alg:jc_work_extraction}
        Require: A sequence of unknown states $\ket{\psi}$
       
        \For {$k=1,2,\ldots$}{
            Receive the unknown state $\ket{\psi}$
            
            Compute direction $\ket{\psi_k}$ using Algorithm~\ref{alg:linucb_vn_var}
            
            Expose it to a field that induces Hamiltonian $H_A=\omega \ketbra{\psi_k}{\psi_k}$. 
            
            Turn on the interaction between the system and the battery whose interaction Hamiltonian is $H_I = \frac{\Omega}{2}(a \otimes \ketbra{\psi_k}{\psi_k^\perp} + a^\dagger \otimes \ketbra{\psi_k^\perp}{\psi_k})$ for a time $t_k=\pi\Omega^{-1}(n_k+1)^{-\frac{1}{2}}$. 
            
            Measure the battery energy to obtain $n_{k+1}$

            Set reward $r_k=\lbrace0,1\rbrace$ according to Eq.~\eqref{eq:reward_jc}
	}
\end{algorithm}

We consider the Jaynes-Cummings work extraction protocol in Algorithm~\ref{alg:jc_work_extraction} which extracts the energy from a field into a battery.

The systems involved in this protocol includes a system in an unknown state $\ket{\psi}$ with a tunable Hamiltonian in the form of $H_A = \nu \ketbra{\phi}{\phi}$ where $\nu$ and $\phi$ can be controlled by the strength and the direction of a field, respectively, as well as a battery system with the Hamiltonian of the battery given by $H_B = \omega a^\dagger a$ where $a=\sum_{n=1}^{\infty} \sqrt{n} \ketbra{n-1}{n}$ and $a^\dagger =\sum_{n=0}^{\infty} \sqrt{n+1} \ketbra{n+1}{n}$. We are also allowed to turn on an interaction $H_I=\frac{\Omega}{2}(a \otimes \ketbra{\psi_k}{\psi_k^\perp} + a^\dagger \otimes \ketbra{\psi_k^\perp}{\psi_k})$ between the system and the battery. 

The protocol works in round $k=1,2,\ldots,N$. 

In each round $k$, we possess a battery state $\ket{n_k}$, receive an unknown state $\ket{\psi}$ and compute a direction $\ket{\psi_k}$ from previous records of $\lbrace n_s\rbrace_{s=1}^{k}$. 

We then expose the system to a field that induces Hamiltonian $H_A=\omega \ketbra{\psi_k}{\psi_k}$, which transfer energy from the field to the system. 

We turn on the interaction $H_I$ between the system and the battery for time $t_k = \pi\Omega^{-1}(n_k+1)^{-\frac{1}{2}}$. The time evolution of the system and the battery is under the total Hamiltonian
\begin{align}
    H= \omega (\ketbra{\psi_k}{\psi_k} + a^\dagger a) + \frac{\Omega}{2}(a \otimes \ketbra{\psi_k}{\psi_k^\perp} + a^\dagger \otimes \ketbra{\psi_k^\perp}{\psi_k}). 
\end{align}
This is exactly given by the famous Jaynes-Cummings model~\cite{Jaynes_1963}, which is nowadays a textbook model~\cite{Scully_1997}. One may verify that the eigenstates of the total Hamiltonian are $\ket{0}\ket{\psi_k^\perp}$ as well as 
\begin{align}
    \ket{n,+}  = \frac{1}{\sqrt{2}}(\ket{n-1}\ket{\psi_k} + \ket{n}\ket{\psi_k^\perp}), \quad 
    \ket{n,-}  = \frac{1}{\sqrt{2}}( \ket{n-1}\ket{\psi_k} - \ket{n}\ket{\psi_k^\perp}),
\end{align}
for $n=1,2,\ldots$ and the eigenvalues are respectively $E_0=0$ and 
\begin{align}
    E_{n+}  =  n\omega + \frac{\Omega}{2} \sqrt{n}, \quad 
    E_{n-}  =  n\omega - \frac{\Omega}{2} \sqrt{n}.
\end{align}
for $n=1,2,\ldots$. The state $\ket{n_k}\ket{\psi}$ is decomposed into $4$ eigenstates $\ket{n_k,\pm}$ and $\ket{n_k+1,\pm}$, i.e.
\begin{align}
    \ket{n_k}\ket{\psi} = \frac{1}{\sqrt{2}}\langle \psi_k^\perp |\psi \rangle(\ket{n_k,+}-\ket{n_k,-}) + \frac{1}{\sqrt{2}} \langle \psi_k |\psi \rangle (\ket{n_k+1,+}+\ket{n_k+1,-}). 
\end{align}
These eigenstates gains phase factors during the time evolution. After time $t_k=\pi\Omega^{-1}(n_k+1)^{-\frac{1}{2}}$, the state evolves to, up to an irrelevant global phase, 
\begin{align}
    e^{-i H t_k}\ket{n_k}\ket{\psi} & = \frac{1}{\sqrt{2}}\langle \psi_k^\perp |\psi \rangle(e^{i\theta_k}\ket{n_k,+}-e^{-i\theta_k}\ket{n_k,-})  + \frac{1}{\sqrt{2}} e^{i\frac{\omega\pi}{\Omega\sqrt{n_k+1}} } \langle \psi_k |\psi \rangle(i\ket{n_k+1,+}-i\ket{n_k+1,-}) \\
    & = \langle \psi_k^\perp |\psi \rangle( i \sin\theta_k \ket{n_k-1}\ket{\psi_k}+ \cos\theta_k \ket{n_k}\ket{\psi_k^\perp}) +  \frac{1}{\sqrt{2}} i e^{i\frac{\omega\pi}{\Omega\sqrt{n_k+1}} } \langle \psi_k |\psi \rangle \ket{n_k+1}\ket{\psi_k^\perp}, 
\end{align}
where $\theta_k = \frac{\pi}{2}\sqrt{\frac{n_k}{n_k+1}}$. 

We finally measure the battery energy. The measurement outcome is $n_{k+1}$. The probability distribution of $n_{k+1}$ is given by 
\begin{align}
    \Pr[n_{k+1}] = \begin{cases}
        |\langle \psi_k |\psi \rangle |^2, \quad & n_{k+1} = n_k+1,\\
        |\langle \psi_k^\perp |\psi \rangle |^2 \cos^2\theta_k, \quad & n_{k+1} = n_k,\\
        |\langle \psi_k^\perp |\psi \rangle |^2 \sin^2\theta_k, \quad& n_{k+1} = n_k-1. 
    \end{cases}
\end{align}
The extracted work is defined as $\Delta W_k =\omega( n_{k+1}-n_k) $. The expected extracted work is given by 
\begin{align}
    \Ex[\Delta W_k] = \omega( |\langle \psi_k |\psi \rangle |^2 - |\langle \psi_k^\perp |\psi \rangle |^2 \sin^2\theta_k )= \omega( |\langle \psi_k |\psi \rangle |^2 (1+\sin^2\theta_k) - \sin^2\theta_k), 
\end{align}
where we have used $|\langle \psi_k^\perp |\psi \rangle |^2 = 1 - |\langle \psi_k |\psi \rangle |^2 $. It is obvious that $\Ex[\Delta W_k]$ increases as $|\langle \psi_k |\psi \rangle |^2$ increases, therefore, $\Ex[\Delta W_k]$ is maximized at $|\langle \psi_k |\psi\rangle|=1$, 
\begin{align}
    \max_{\ket{\psi_k}} \Ex[\Delta W_k] = \omega. 
\end{align}
The dissipation in this round is thus given by the difference between the maximal and the actual expected extracted work
\begin{align}
    \dissipation^{\text{jc},k} = \max_{\ket{\psi_k}} \Ex[\Delta W_k] -  \Ex[\Delta W_k] = \omega (1+\sin^2\theta_k^2)(1- |\langle \psi_k|\psi\rangle |^2 ) \leq 2 \omega (1-|\langle \psi_k|\psi\rangle |^2). 
\end{align}
The correspondence between the reward measurement and the battery energy measurement is given by
\begin{align}
    \label{eq:reward_jc}
    r_k = \begin{cases}
        1, \quad & n_{k+1} = n_k+1, \\
        0, \quad & \text{otherwise. }
    \end{cases}
\end{align}
The probability distribution of the reward is 
\begin{align}
    \Pr[R_k=r_k] = \begin{cases}
        |\langle \psi_k |\psi \rangle |^2, \quad & r_k=1,\\
        |\langle \psi_k^\perp |\psi \rangle |^2 , \quad & r_k=0.
    \end{cases}
\end{align}
The regret in this round is thus 
\begin{align}
    \regret^k = 1-|\langle \psi_k |\psi \rangle |^2. 
\end{align}
The dissipation and the regret in this round is thus related by 
\begin{align}
    \dissipation^{\text{jc},k} \leq 2\omega \regret^k .
\end{align}
The cumulative dissipation over $N$ rounds is thus 
\begin{align}
    \dissipation^{\text{jc}}(N)  = \sum_{k=1}^{N}  \dissipation^{\text{jc},k} \leq 2 \omega\sum_{k=1}^{N} (1-|\langle\psi_k|\psi\rangle|^2). 
\end{align}
Which is related to the cumulative dissipation in extracting work from knowledge via 
\begin{align}
    \dissipation^{\text{jc}}(N)\leq 2\omega \regret (N), 
\end{align}
\begin{theorem} There exists an explicit protocol for extracting work from knowledge that adaptively updates the estimate $\ket{\hat{\psi}_k}$ and the probe state $\ket{\psi_k}$ based on the rewards $\lbrace r_s \rbrace_{s=1}^{k-1}$, achieving, with probability at least $1-\delta$, 
\begin{align} 
\dissipation^{\text{jc}} (N) = O\left( \omega \ln (N) \ln \left( \frac{N}{\delta} \right)  \right). 
\end{align} 
\end{theorem}

\section{Cost of measurement and erasure}
\label{sec:landauer}
The measurement as well as erasure of memory of the agent does not come for free. The cost of measurement along with the cost of memory erasure can be lower bounded by a quantity known as \emph{QC-mutual
information}, $I_{QC}$, which is a measure of how much information a measurement carries about the quantum system \cite{Sagawa_2009}. $I_{QC}$ is then upper bounded by the entropy of the classical memory register itself. Hence the total cost of measurement and erasure can be loosely lower bounded by just the entropy of the memory register itself.
By Landauer's principle~\cite{Sagawa_2009,Goold_2015,Tan_2022}, the heat dissipation required to erase the register is lower bounded by the entropy change $\Delta S$ of the memory register via
\begin{align}
    \beta Q \geq \Delta S.
\end{align}
Furthermore, it is widely accepted that the lower bound can be achieved in the quasi-static limit when the state to be erased is known~\cite{riechers2021initial,Jun_2014,Miller_2020}. In our work extraction model, we repeatedly perform measurements in the energy eigenbasis, which requires memory erasure in order to store new measurement outcomes. With the assumption on the probability distribution of work values scaling linearly with fidelity, when $\ket{\psi}$ is known, the measurement outcome is deterministic and there is no energy dissipation. However, when $\ket{\psi}$ is unknown, the measurement outcome is stochastic and there is energy dissipation. This may also be taken into account in the dissipation, that is, 
\begin{align}
    \label{eq:regret_modified}
    \dissipation'(N) = \dissipation(N) + \beta^{-1} \sum_{k=1}^N  \Delta S_k, 
\end{align}
where $\Delta S_k$ is the entropy change of the memory register in round $k$. Let $\epsilon_k = 1 - |\langle \psi_k | \psi\rangle |^2$ be the infidelity in round $k$. $\Delta S_k$ is closely related to $\epsilon_k$ in both models we consider. In the semi-classical battery model, the entropy change is upper bounded by 
\begin{align}
    \Delta S_k = -(1-\epsilon_k) \ln (1-\epsilon_k) - \epsilon_k \ln \epsilon_k \leq  \epsilon_k - \epsilon_k \ln \epsilon_k, 
\end{align}
and the dissipation is upper bounded by 
\begin{align}
    \dissipation^{\text{sc},*}(N) = \dissipation^{\text{sc}}(N) + \beta^{-1} \sum_{k=1}^N (\epsilon_k- \epsilon_k\ln\epsilon_k). 
\end{align}
In the Jaynes-Cummings battery model, the entropy change in round $k$ is upper bounded by 
\begin{align}
    \Delta S_k = -(1-\epsilon_k) \ln (1-\epsilon_k) - \epsilon_k \ln \epsilon_k - \epsilon_k (\cos^2\theta_k \ln \cos^2\theta_k + \sin^2\theta_k \ln \sin^2\theta_k ) \leq 2\epsilon_k -\epsilon_k\ln \epsilon_k, 
\end{align}
and the dissipation is upper bounded by 
\begin{align}
    \dissipation^{\text{jc},*}(N) = \dissipation^{\text{jc}}(N) + \beta^{-1} \sum_{k=1}^N (2\epsilon_k - \epsilon_k\ln\epsilon_k). 
\end{align}

The quantum state tomography algorithm in~\cite{lumbreras24pure} ensures a polylogarithmic cumulative infidelity with a high probability. Now we focus on the case where the cumulative infidelity is upper bounded by $\ln\frac{N}{\delta}\ln N$ with high probability $1-\delta$, that is, for some constant $C$, 
\begin{align}
    \sum_{k=1}^N \epsilon_k \leq C \ln \frac{N}{\delta}\ln N.
\end{align}
By taking the derivative $(-\epsilon_k\ln\epsilon_k)' = -  \ln\epsilon_k - 1$, it is obvious that when $0<\epsilon_k \leq e^{-1}$, $-\epsilon_k\ln  \epsilon_k$ increases with $\epsilon_k$ while $e^{-1}\leq \epsilon_k\leq 1$, $-\epsilon_k\ln\epsilon_k$ decreases with $\epsilon_k$. Now we consider two cases: 

\textbf{Case 1: $C (\ln (N/\delta) \ln N)/N \leq e^{-1}$.} This happens when $N$ is large enough. As a result, $-\epsilon_k\ln\epsilon_k$ monotonically increase with respect to $\epsilon_k$ when $\epsilon_k\leq C  (\ln (N/\delta) \ln N)/N$. On the one hand, for $\epsilon_k\leq  C(\ln (N/\delta) \ln N)/N $, 
\begin{align}
    -\epsilon_k\ln\epsilon_k\leq \frac{C\ln\frac{N}{\delta} \ln N}{N} \ln \frac{N}{C\ln\frac{N}{\delta} \ln N},
\end{align}
On the other hand, for $\epsilon_k > C (\ln (N/\delta) \ln N)/N$, 
\begin{align}
    -\epsilon_k\ln\epsilon_k \leq  \epsilon_k \ln\frac{N}{C\ln \frac{N}{\delta} \ln N}, 
\end{align}
Then, 
\begin{align}
    -\sum_{k=1}^{N} \epsilon_k \ln\epsilon_k & = - \sum_{\epsilon_k:\epsilon_k\leq \frac{C \ln \frac{N}{\delta} \ln N}{N}} \epsilon_k \ln\epsilon_k - \sum_{\epsilon_k:\epsilon_k >\frac{C \ln \frac{N}{\delta} \ln N}{N}}\epsilon_k \ln\epsilon_k \\
    & \leq \sum_{\epsilon_k:\epsilon_k\leq \frac{C \ln \frac{N}{\delta} \ln N}{N}} \frac{C \ln \frac{N}{\delta} \ln N}{N} \ln \frac{N}{C \ln \frac{N}{\delta} \ln N}+ \sum_{\epsilon_k:\epsilon_k>\frac{C \ln \frac{N}{\delta} \ln N}{N}} \epsilon_k  \ln\frac{N}{C \ln \frac{N}{\delta} \ln N}  \\
    & \leq  C \ln \frac{N}{\delta} \ln N \ln \frac{N}{C \ln \frac{N}{\delta} \ln N} + C \ln \frac{N}{\delta} \ln N \ln \frac{N}{C \ln \frac{N}{\delta} \ln N} \leq 2 C \ln \frac{N}{\delta} \ln N \ln \frac{N}{C \ln \frac{N}{\delta} \ln N}. 
\end{align}

\textbf{Case 2: $C (\ln (N/\delta) \ln N)/N > e^{-1}$.} This happens when $N$ is not very large. Therefore, $N/e\leq  C \ln (N/\delta) \ln N$ and certainly $N \geq 2$. Using the fact that $-\epsilon_k\ln\epsilon_k\leq e^{-1}$, we obtain
\begin{align}
    \sum_{t=1}^{N} \epsilon_k \ln \frac{1}{\epsilon_k} \leq \frac{N}{e} \leq C \ln \frac{N}{\delta} \ln N . 
\end{align}
Combining both cases, we conclude that 
\begin{align}
    - \sum_{t=1}^{N} \epsilon_k \ln\epsilon_k \leq O((\ln N)^3). 
\end{align}
Substituting into Eq.~\eqref{eq:regret_modified}, we obtain the dissipation for for the semi-classical model,  
\begin{align}
    \dissipation^{\text{sc},*}(N)  \leq  \dissipation^{\text{sc}}(N) + \beta^{-1} O((\ln N)^3) = O((\ln N)^3). 
\end{align}
and the Jaynes-Cummings battery model, 
\begin{align}
    \dissipation^{\text{jc},*}(N)  \leq  \dissipation^{\text{jc}}(N) + 2\beta^{-1} O((\ln N)^3) = O((\ln N)^3), 
\end{align}
Therefore, the regret still scales as $O(\polylog(N))$ even if we take the energy dissipation due to Landauer's principle into account. 

As mentioned above, in order to achieve Landauer's bound, the erasure process must necessarily be done quasi-statically. This is not something that a sequential adaptive agent can accomplish since it needs to measure and decide on its action in real time. To circumvent this, an array of empty memory registers can be first prepared. Suppose the agent will be extracting work for $N$ steps, we first prepare a memory register $M_0$, consisting of $N$ empty registers.
\begin{equation}
    M_0 = \{\underbrace{0,0,\ldots,0}_N\}
\end{equation}
At every time step, rather than erasing the old memory that has the previous outcome remembered, the agent simply records the outcome of the measurement on the battery into a new empty register. In order words, after time step $t$, the memory register takes the form
\begin{equation}
    M_t = \{r_1,r_2,\ldots,r_t,0,\ldots,0\} \quad \text{for}\quad t\in \lbrace{1,2,\ldots N \rbrace}~.
\end{equation}
At the end of all the extractions, the memory $M_N$ can then be quasi-statically reset. As previously calculated, as $t\to N$, the distribution of $r_t$ becomes more peaked and the total entropy of the memory registers scale with $O(\ln N)^3$, likewise for the cost of measurement. The resultant dissipation therefore still scales with $O(\polylog(N))$.

\end{document}